\documentclass[letterpaper,10pt]{extarticle}  
\usepackage{import}
\usepackage{local}
\usepackage[left=.65in,top=.65in,bottom=.65in,right=.65in]{geometry}
\usepackage{listings}
\definecolor{codegreen}{rgb}{0,0.6,0}
\definecolor{codegray}{rgb}{0.5,0.5,0.5}
\definecolor{codepurple}{rgb}{0.58,0,0.82}

\usepackage{lipsum} 

\title{BHiCect 2.0: Multi-resolution clustering of Hi-C data}

\author{%
\textbf{Vipin Kumar\textcolor{Accent}{\textsuperscript{1,*}}, %
Roberto Rossini\textcolor{Accent}{\textsuperscript{2}}, %
Jonas Paulsen\orcidlink{0000-0002-7918-5495}\textcolor{Accent}{\textsuperscript{2}}, %
and Anthony Mathelier\orcidlink{0000-0001-5127-5459}\textcolor{Accent}{\textsuperscript{1,3,4}}}\\
\begin{small}\textcolor{Accent}{\textsuperscript{1}}Norwegian Centre for Molecular Biosciences and Medicine (NCMBM), Nordic EMBL Partnership, University of Oslo, 0318 Oslo, Norway\\
\textsuperscript{2}Department of Biosciences, University of Oslo, 0316, Oslo, Norway\\
\textsuperscript{3}Bioinformatics in Life Science (BiLS) initiative, Department of Pharmacy,  University of Oslo, Oslo, Norway\\
\textsuperscript{4}Department of Medical Genetics, Institute of Clinical Medicine, University of Oslo and Oslo University Hospital, Oslo, Norway\\ 
\textcolor{Accent}{\textsuperscript{*}}Correspondence: \textcolor{Accent}{vipin.kumar@ncmbm.uio.no} \\ \end{small}
}

\date{}
\begin{document}
\maketitle
\thispagestyle{empty}

\section{Abstract}

\begin{doublespacing}

\noindent
\textbf{\textcolor{Accent}{Chromatin conformation capture technologies such as Hi-C have revealed that the genome is organized in a hierarchy of structures spanning multiple scales observed at different resolutions. Current algorithms often focus on specific interaction patterns found at a specific Hi-C resolution. We present BHi-Cect 2.0, a method that leverages Hi-C data at multiple resolutions to describe chromosome architecture as nested preferentially self-interacting clusters using spectral clustering. This new version describes the hierarchical configuration of chromosomes by now integrating multiple Hi-C data resolutions. Our new implementation offers a more comprehensive description of the multi-scale architecture of the chromosomes. We further provide these functionalities as an R package to assist their integration with other computational pipelines.\\
The BHiCect 2.0 R packages is available on github at \url{https://github.com/princeps091-binf/BHiCect2} with the version used for this manuscript on Zenodo at \url{https://doi.org/10.5281/zenodo.17985844}.
}}

\section{Introduction}
Conventional methods describing chromatin architecture in the context of Hi-C aim to highlight specific interaction patterns expected to reflect particular chromosome structures. For instance, topologically associating domains (TADs) \cite{Dixon2012-ao,Nora2012-so,Hou2012-bv,Sexton2012-wj} are derived from diagonal blocks and compartments from plaid patterns; they represent outstanding illustrations of this approach \cite{Lieberman2009}. Importantly, these pattern-specific approaches do not produce a description of the chromosomes' conformation at multiple resolutions and scales, potentially leaving important patterns undetected.

To address the limitation of pattern-specific methods, we previously developed BHiCect \cite{Kumar2020}, which detects preferential self-interactions (i.e., chromatin regions interacting more frequently with themselves than with others) across scales to more comprehensively detect the multitude of chromatin structures, such as loops, TADs, compartments, and higher-order aggregates that constitute the chromatin architecture.

Notably, the original implementation of BHiCect relied on a specific resolution of the Hi-C data - i.e., the bin sizes of the HI-C data - preselected by the user. This setting imposed a particular limitation, as different resolutions allow for highlighting different scales of interaction patterns. Specifically, high-resolution Hi-C data highlight detailed short-range interaction patterns, while low-resolution data comprehensively capture long-range interaction patterns \cite{LAJOIE201565}.

Moreover, other conventional methods that cluster Hi-C data at a single predefined resolution do not address the differential coverage of short (high-resolution) and long-range (low-resolution) interactions across Hi-C data resolutions. To leverage the complementarity of high- and low-resolution Hi-C data and to address the single-resolution limitation of BHiCect, we updated the framework to perform clustering of chromatin interactions by dynamically selecting the highest resolution that maximises the coverage of the interactions captured by HiC for each cluster. Formally, BHiCect outputs a binary tree that provides a comprehensive, multi-resolution decomposition of chromatin interactions into clusters of preferential self-interactions. This representation facilitates the exploration of how multi-scale chromatin interactions are hierarchically nested and interact with one another.

\begin{figure*}[tbh]%
\centering
{\includegraphics[width=\textwidth]{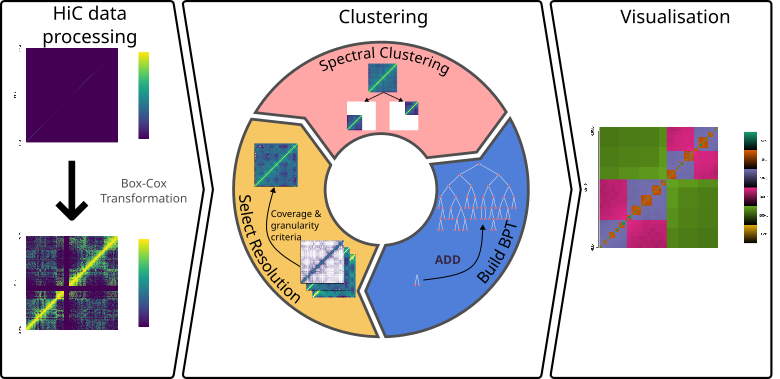}}
\caption{\textbf{Workflow of BHiCect 2.0}. BHiCect 2.0 consists of three main phases. In the first step, BHiCect 2.0 preprocesses the normalised HiC data using Box-Cox transformation (left). It is followed by performing the spectral clustering at selected resolutions to build the bipartite tree (BPT) (middle). BHiCect 2.0 finally provides visualisations of the clusters of interest (right).}\label{fig1}
\end{figure*}

\section{Methods and implementation}\label{sec2}

Like the original version, BHiCect 2.0 relies on spectral clustering to recursively bi-partition the genome (represented as a weighted graph with corresponding Hi-C signal as edge weights between Hi-C bins) into preferentially self-interacting clusters. High-resolution Hi-C data capture short-range chromatin interactions (e.g., loops and sub-TADs), whereas lower resolutions capture long-range chromatin interactions (e.g., compartments and higher-order aggregates). BHiCect 2.0 automatically determines the optimal Hi-C data resolution at each iteration of the spectral clustering process. This approach ensures that both short- and long-distance interaction patterns are accurately represented throughout the binary tree. Our selection criteria identify the finest Hi-C data resolution for which 99\% of the interaction matrix is non-empty. Another constraint is to enforce that the selected bin size (i.e., resolution) remains constant or decreases at every iteration of the spectral clustering.

\subsection{BHi-Cect 2.0 R package}\label{subsec1}

BHiCect 2.0 is implemented as an R package available online (https://github.com/princeps091-binf/BHiCect2) with a BHiCect function that accepts Hi-C data and a user-defined set of resolutions to apply the spectral clustering algorithm. The package enables parallel computation to leverage the available computing resources. Moreover, the package is accompanied by the BHiCectViz helper package, which implements heatmap routines to visually interpret the BHiCect 2.0 binary tree output. As illustrated in Figure~\ref{fig1}, a typical use case would consist of:

\begin{itemize}
    \item running the BHiCect function to perform the multi-resolution spectral clustering
    \item visualising the clustering results using the Heatmap visualisation function on the desired cluster
    \item diagnosing the cluster of interest
\end{itemize}

\begin{lstlisting}[caption= Typical usecase code, language=R]{Code}

res_set <- c('1Mb','500kb','100kb','50kb','10kb','5kb')
res_num <- c(1e6L,5e5L,1e5L,5e4L,1e4L,5e3L)
names(res_num) <- res_set
BHiCect_results <- BHiCect(label = res_set,
                         bin_sizes = res_num,
                         data = chr_dat_l,
                         nworkers = 4)

\end{lstlisting}

\subsection{Recovery of known structures (TADs)}\label{subsec2}
We examined how BHiCect 2.0 recovers TADs, as they constitute one of the outstanding results from clustering methods applied to Hi-C \cite{Dixon2012-ao,Nora2012-so,Hou2012-bv,Sexton2012-wj}. The exact definition of a TAD remains an active field of study. It has prompted the development of various methods, each aiming to focus on a specific form of TADs characterised by a particular interaction pattern. 
Here, we aim to demonstrate how BHiCect 2.0 can reliably recover various TADs, as identified by multiple methods. This illustrates how our pattern-agnostic approach implicitly incorporates a variety of structures while maintaining consistency with a hierarchically nested configuration for the overall chromosome architecture.

\begin{figure*}%
\centering
{\includegraphics[width=.5\textwidth]{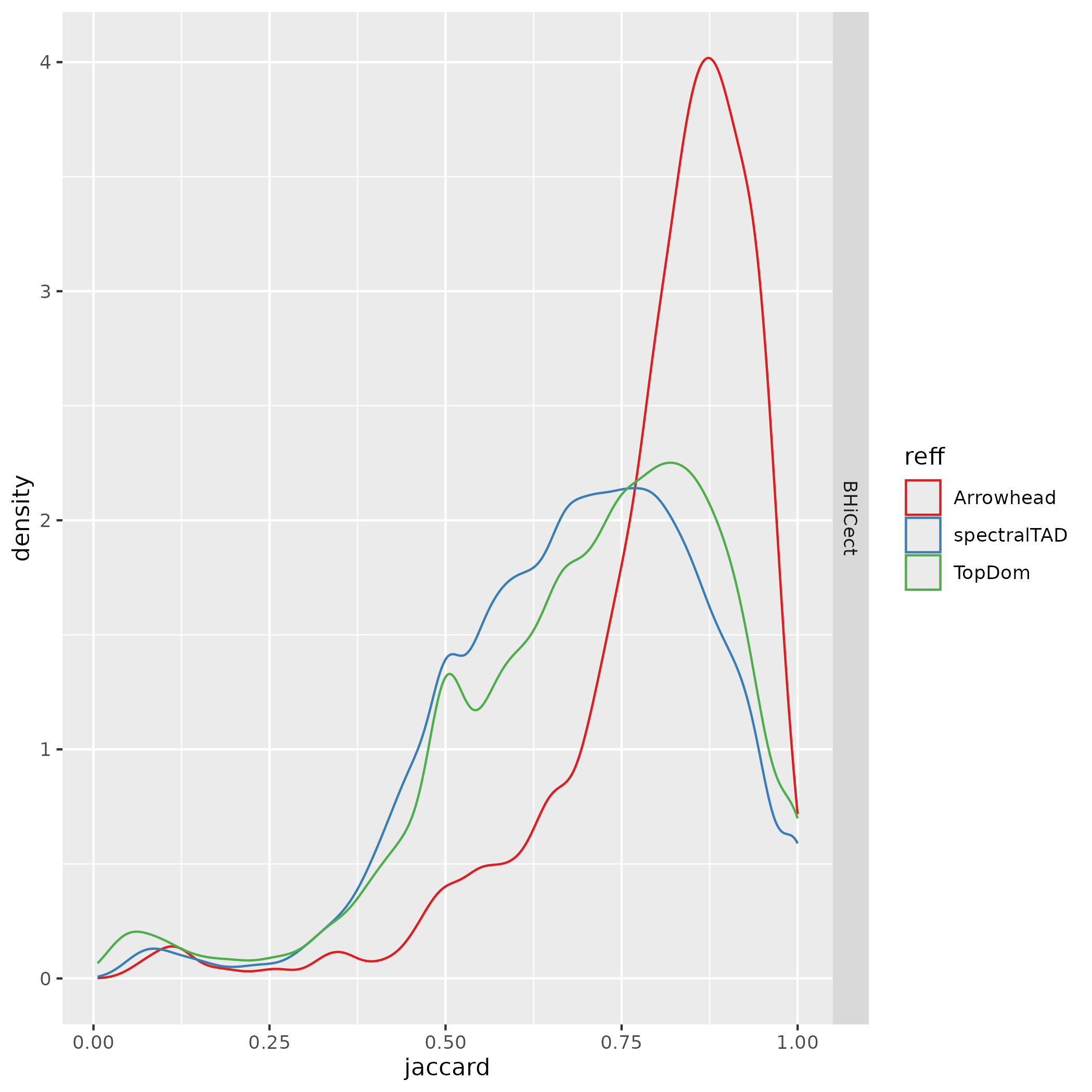}}
\caption{\textbf{BHiCect captures TADs from multiple callers}. Density plot of Jaccard indices computed for every TAD found by the method indicated by the colour code and the best matching BHiCect 2.0 cluster.}\label{fig2}
\end{figure*}

We quantify the recovery of TADs across two methods using the Jaccard index \cite{Favorov2012-eo}. Given the lack of gold standards regarding effective/functional TADs, we must examine how various TAD-calling methods agree. We do so by ”alternating” the method whose set of TADs we consider as the reference against which the other methods' results must agree. We considered Arrowhead \cite{Rao2014}, SpectralTAD \cite{spectralTAD}, and TopDom \cite{topdom} as they rely on different assumptions and methodologies to call TADs. Each method was ran genome-wide on the Rao et al. dataset for GM12878 at 10kb resolution \cite{Rao2014}. We represent the level of agreement/recovery of a particular method for a specific reference TAD set as a distribution of Jaccard indices (each TAD from the reference set is mapped to the best matching TAD produced by the considered method based on their Jaccard index). We thus obtain a collection of Jaccard index distributions corresponding to their capacity to recover the TADs from the other reference methods with which they were compared. Given that BHiCect 2.0 is a multi-resolution method, expecting other "single-resolution" methods to recover the resulting clusters would be inadequate.  We, therefore, do not consider it a reference set in this analysis. The distribution of Jaccard index scores indicates that, in most cases, BHiCect 2.0 reliably recovers the TADs detected by other methods with values above 0.5 (Figure~\ref{fig2}). As a comparison, the other methods highlighted more discrepancies between the called TADs with Arrowhead poorly recovering the TADs produced by spectralTAD and TopDom (peak in density around 0 in Supplementary Figure~1), while SpectralTAD and TopDom and exhibit both poorly and well-recovered TADs (see the bi-modal distributions in Supplementary Figure 2 and 3).

In conclusion, BHiCect 2.0 can faithfully reproduce TADs from multiple methods, as indicated by the agreement with BHiCect clusters using the Jaccard index. However, the multi-resolution clustering enables a finer recovery of these canonical structures without requiring the user to specify a single specific predefined resolution.

\section{Conclusion}

We provide BHiCect 2.0 to produce a more concrete representation of the multi-scale chromosome architecture captured by Hi-C. More specifically, the explicit incorporation of multiple resolutions offers a richer and more faithful description of the chromosome’s hierarchical structure. The coverage criteria used by BHiCect 2.0 to identify the adequate resolution for spectral clustering directly considers the experimental limitation of sequencing depth, which determines the effective resolutions of Hi-C. Thus, this approach aims to systematically and accurately extract the interaction signal from Hi-C data at all scales by considering the optimal resolution at each step of the decomposition.

The addition of original clustering diagnostics and the helper package for visualization offers a valuable complement for better interpreting the results and exploring the clusters detected further.
On the computational side, this new implementation now offers better tooling and parallel computation options that better leverage the resources of modern machines.

As BHiCect produces nested clusters at multiple resolutions, we acknowledge that it might be favored when evaluating TAD recovery. Nevertheless, BHiCect was not designed to specifically detect TADs, but its hierarchical and multi-resolution clustering approach enabled an adaptive and comprehensive description of chromosome architecture that incorporates multiple TAD definitions detected by different methods. This opens up the possibility of examining the structural relation with different kinds of structure within the chromosome and characterise the broader architectural context of TADs.

More broadly, our ‘pattern-agnostic’ strategy incorporates a variety of patterns beyond the typical diagonal blocks or plaid. The benefit of this greater pattern variety is supported by the growing evidence of the importance of novel patterns, such as strips, to infer mechanistic insights such as different forms of loop-extrusion \cite{Fudenberg2017-pn}.

\section{Competing interests}
No competing interest is declared.

\section{Author contributions statement}
V.K.: Conceptualization, Formal analysis, Methodology, Software, Investigation, Validation, Visualization, Project administration, Funding acquisition, Writing – original draft, Writing – review and editing. A.M.: Conceptualization, Methodology, Funding acquisition, Writing – review and editing, Project administration, Resources, Supervision. R.R.: Conceptualization, Writing – review and editing. J.P.: Conceptualization, Writing – review and editing.

\section{Acknowledgements}
This word was funded by The Research Council of Norway [187615]; Helse Sør-Øst, and the University of Oslo through the Centre for Molecular Medicine Norway (NCMM) to Mathelier group; Norwegian Cancer Society [215027] to Mathelier group; the European Union’s Horizon 2020 research and innovation program under the Marie Sklodowska-Curie grant agreement [801133 to V.K.]. We thank the NCMBM IT team for IT support, Ingrid Kjelsvik for administrative support, the Kuijjer and Mathelier groups’ members for insightful discussions.

\end{doublespacing}

\renewcommand\refname{References}
\begin{footnotesize}
\bibliographystyle{unsrt.bst} 
\textnormal{\bibliography{reference.bib}}
\end{footnotesize}

\begin{figure}[h!] 
    \centering 
    \includegraphics[width=0.8\textwidth, keepaspectratio]{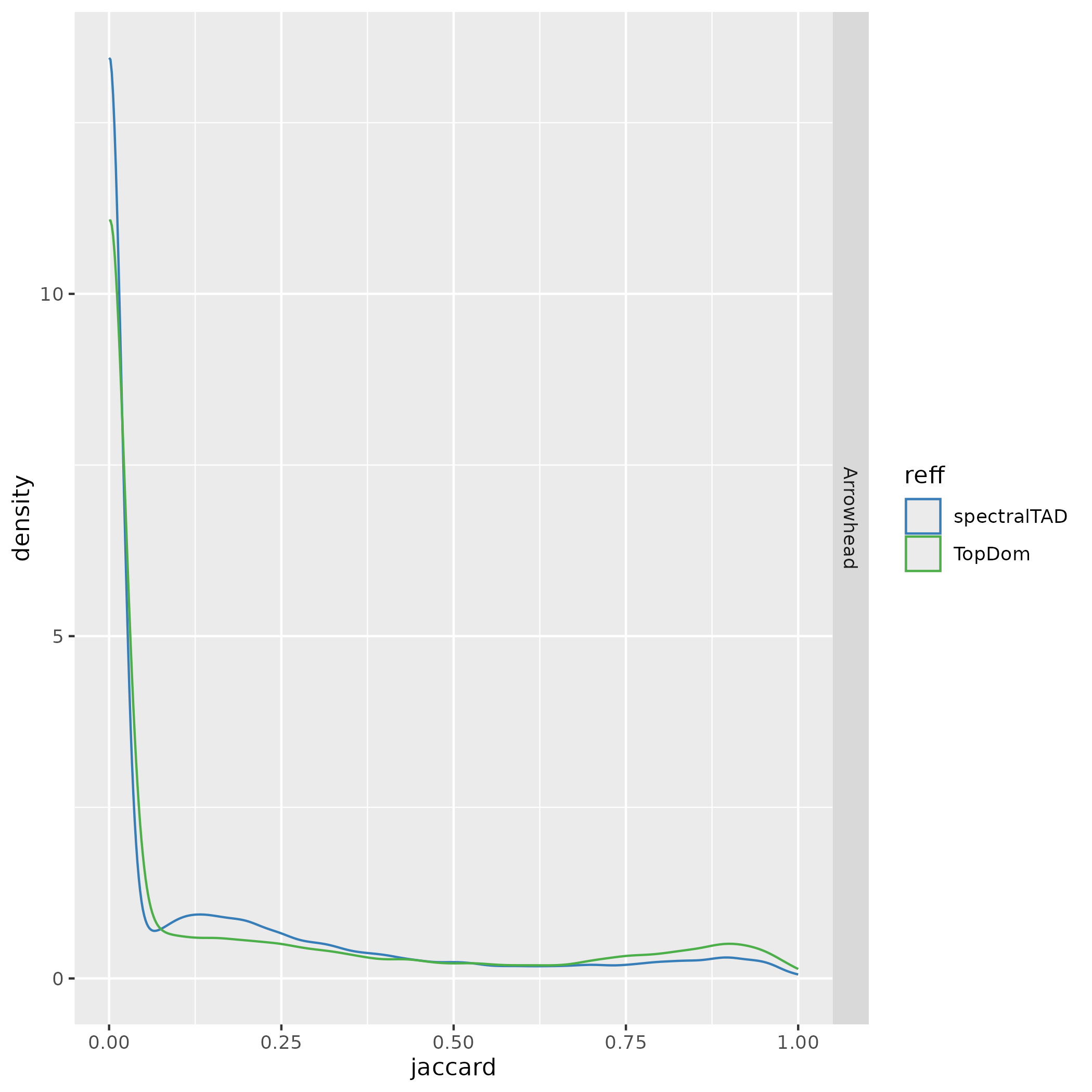} 
    \caption*{
        \textbf{Supplementary Figure 1: Jaccard Index distribution of best matching Arrowhead TADs}
        Density plot for the distribution of Jaccard indices computed for each TAD from the labeled method and their best matching Arrowhead TAD.
    }
    \label{fig:supp_figure1} 
\end{figure}

\clearpage 

\begin{figure}[h!]
    \centering
    \includegraphics[width=0.7\textwidth, keepaspectratio]{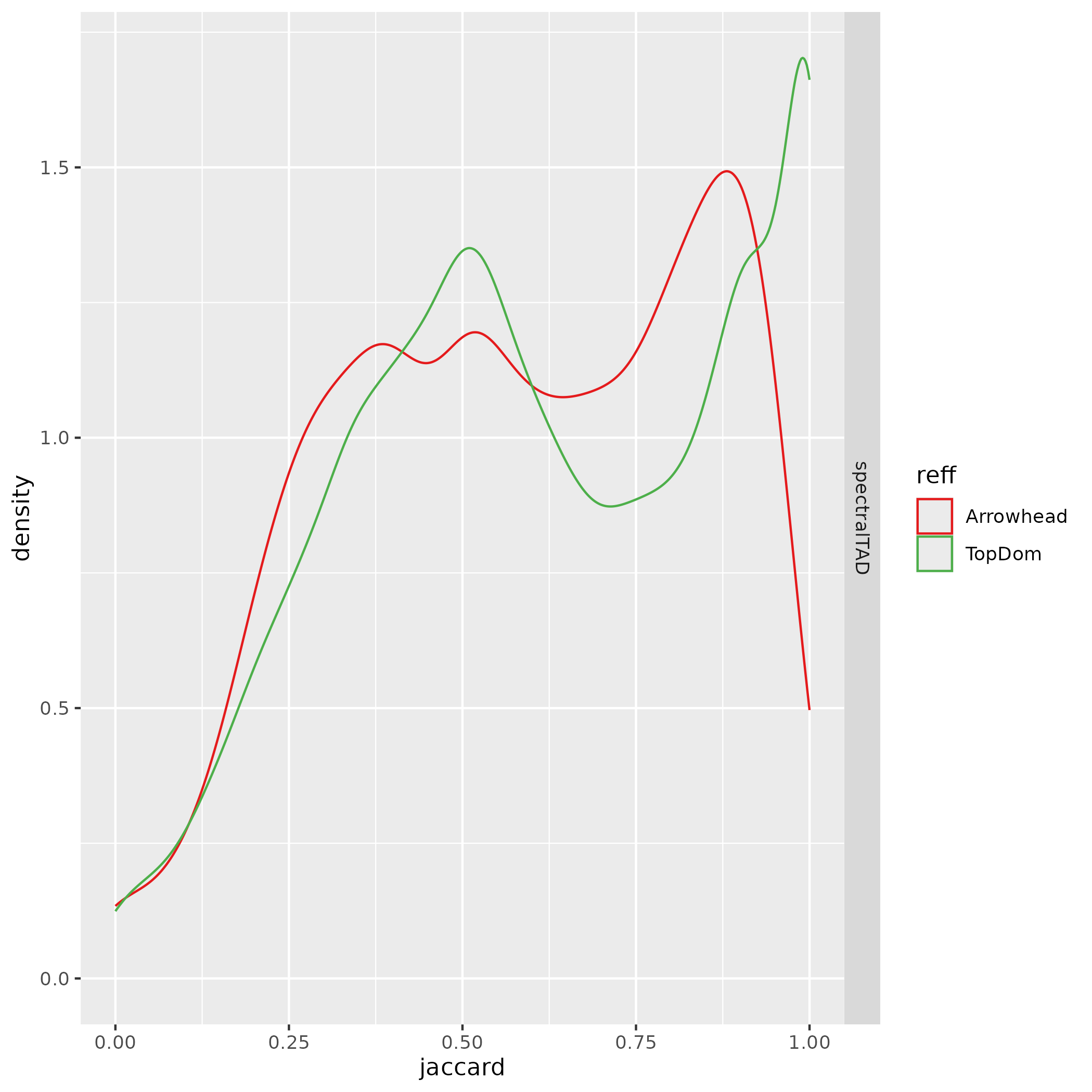} 
    \caption*{
        \textbf{Supplementary Figure 2: Jaccard Index distribution of best matching spectralTAD TADs}
        Density plot for the distribution of Jaccard indices computed for each TAD from the labeled method and their best matching spectralTAD TAD.
    }
    \label{fig:supp_figure2}
\end{figure}

\begin{figure}[h!]
    \centering
    \includegraphics[width=0.7\textwidth, keepaspectratio]{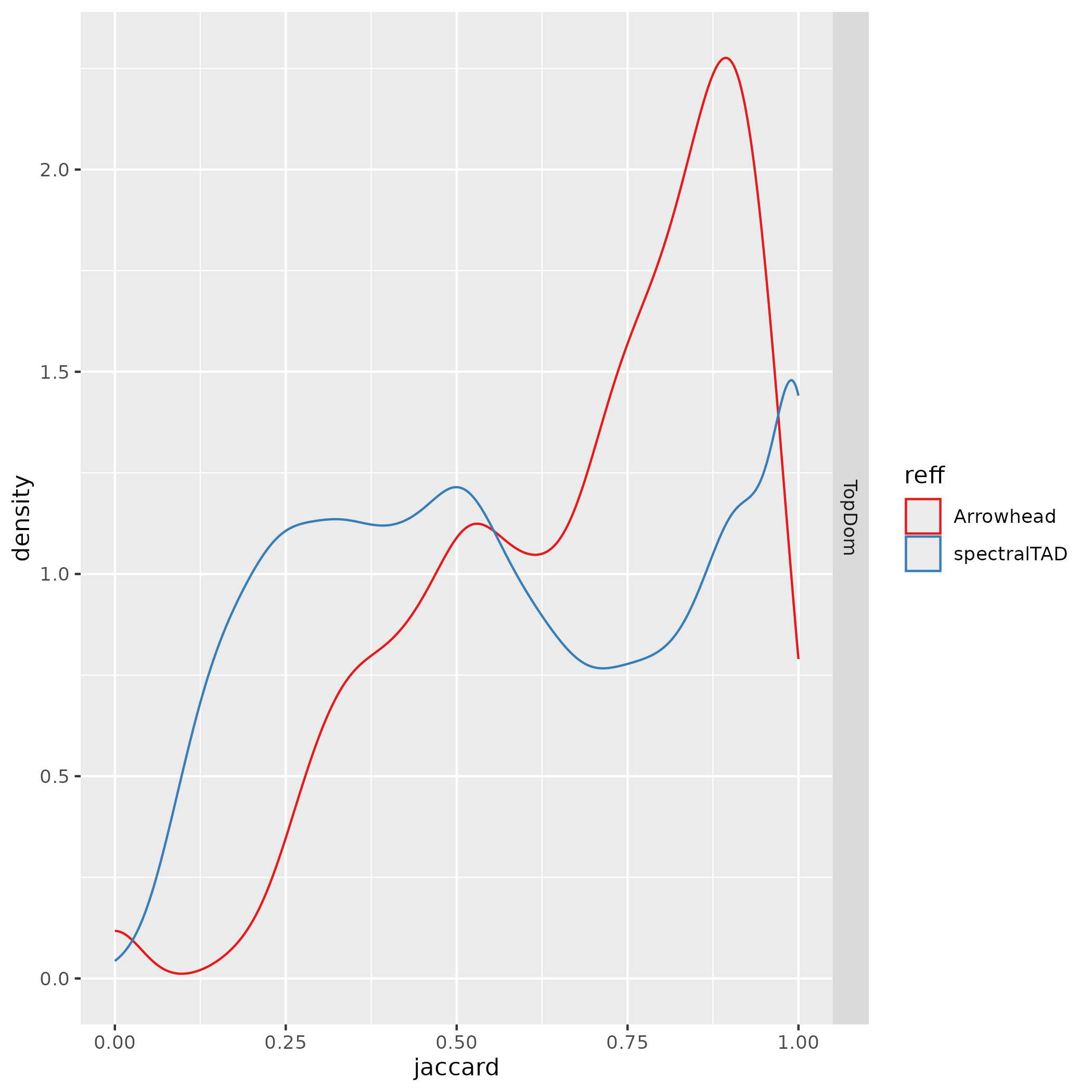} 
    \caption*{
        \textbf{Supplementary Figure 3: Jaccard Index distribution of best matching TopDom TADs}
        Density plot for the distribution of Jaccard indices computed for each TAD from the labeled method and their best matching TopDom TAD.
    }
    \label{fig:supp_figure3}
\end{figure}

\end{document}